\documentclass[12pt]{article}\usepackage{amssymb}

\usepackage{graphicx}
\usepackage{longtable}

\begin{document}
\title{Chemical Potential Dependence of Chiral Quark Condensate in
Dyson-Schwinger Equation Approach of QCD}

\author{{Lei Chang$^{1}$, Huan Chen$^{1}$, Bin Wang$^{1}$, Wei Yuan$^{1}$,
and Yu-xin Liu$^{1,2,3,}$\thanks{corresponding author} }\\[3mm]
\normalsize{$^1$ Department of Physics and MOE Key Laboratory of
Heavy Ion Physics,}\\
\normalsize{Peking University, Beijing 100871, China}\\
\normalsize{$^2$ CCAST (World Laboratory), P.O. Box 8730, Beijing
100080, China} \\
\normalsize{$^3$ Center of Theoretical Nuclear Physics, National
Laboratory of Heavy Ion Accelerator,}\\ \normalsize{ Lanzhou 730000,
China}  }

\maketitle


\begin{abstract}
We propose a chemical potential dependent effective gluon
propagator and study the chiral quark condensate in strongly
interacting matter in the framework of Dyson-Schwinger equation
approach. The obtained results manifest that, as the effect of the
chemical potential on the effective gluon propagator is taken into
account, the chiral quark condensate decreases gradually with the
increasing of the chemical potential if it is less than the
critical value, and the condensate vanishes suddenly at the
critical chemical potential. The inclusion of the chemical
potential in the effective gluon propagator enhances the
decreasing rate and decreases the critical chemical potential. It
indicates that the chiral symmetry can be restored completely at a
critical chemical potential and restored partially as the chemical
potential is less than the critical value. If the effective gluon
propagator is independent of the chemical potential, the chiral
symmetry can only be restored suddenly but no gradual restoration.

\end{abstract}




\newpage

\parindent=20pt


The chiral symmetry spontaneous breaking and color confinement
have been known as two essential characters of strong interaction
in low energy region. The relation between these two essential
properties is a fundamental problem in strong interaction physics.
However, it is still very difficult to study such a relation
directly at present. On the other hand, it is fortunate that one
has known the vacuum of strong interaction would become trivial at
high enough temperature and/or chemical potential because of the
very weak-coupling interaction, or in other word, the asymptotic
freedom. Then the behaviors of the chiral symmetry restoration and
color deconfinement at hight temperature and/or chemical potential
would give signatures or information about the relation between
the two properties. Concerning the measurement of the chiral
symmetry restoration, one used to take the chiral quark condensate
$\langle \bar{q} q \rangle$ as a characteristic since it is
believed to be one of the most important configuration of the
strong interaction vacuum (see for example Refs.\cite{SS98,BR02}).
As for the methodology, the Lattice QCD simulation can be safely
expand to finite temperature\cite{MST04,Choe01} and has made great
efforts to deal with the problem at finite chemical
potential\cite{FK02,Allton02,Delia03,GG03,Focrand,ADGL05,AFHL05}.
With the difference between the fundamental and adjoint
representations of the color symmetry of a quark being taken into
account, lattice QCD simulation and SU(3) gauge theory analysis
show that the quark deconfinement and chiral symmetry restoration
can happen at same critical temperature or not\cite{MST04,KL99}.
Meanwhile many approaches in the framework of continuous field
theory of QCD have been implemented to study the chiral symmetry
breaking and its restoration in dense strongly interacting matter
at zero temperature, and different models gave different behaviors
at finite chemical potential (see for example
Refs.\cite{BC90,CFG92,LK94,Cel95,AB01,MD97,Mits97,
MRS98,CD99,GYL01,Choe02,LGG03,Druk03,Zong05}). However, there is
still not any work to take into account the effect of the chemical
potential dependence of the effective gluon propagator.

It has been shown that the Dyson-Schwinger (D-S) equation approach
provides a nonperturbative framework which admits a simultaneous
study of dynamical chiral symmetry breaking and color
confinement\cite{Craig01,Alkofer01}. With nonperturbative
truncations preserving the chiral symmetry, this approach has been
widely used to study the properties of strong interaction vacuum
and the properties of hadrons in free
space\cite{Craig02,Watson01}. Meanwhile it has been extended to
the system with finite temperature and/or finite chemical
potential to simulate the chiral symmetry restoration and
deconfinement\cite{Harada99,Craig03}. In this paper we will then
take the Dyson-Schwinger equation approach at finite chemical
potential to calculate the chemical potential dependence of the
chiral quark condensate in strongly interacting matter. In
previous studies on the chiral behavior at finite chemical
potential, the effective gluon propagator is usually taken as the
same as that in free space. We should know that the chemical
potential does infect the interaction. Then the chemical potential
dependence of the gluon propagator has to be deliberated. In this
paper we propose an approximation to take into account the
chemical potential dependence of the effective gluon propagator
and investigate the effect of the interaction strength and the
screening parameters in the modified effective gluon propagator on
the chiral quark condensate at finite chemical potential.


As the lowest dimension condensate of quarks and gluons, the chiral
quark condensate is the essential characteristic in the QCD phase
transition, and hence plays essential role in describing hadron
structure and properties of nuclear matter and finite nuclei. In the
chiral limit, the chiral quark condensate can be evaluated from the
quark mass function in ultraviolet region at zero chemical potential
or with the definition
\begin{equation}  \label{Qcond0mu}
|\langle\bar{q}q\rangle| = Tr_{D,C} \int
\frac{d^{4}q}{(2\pi)^{4}}G(q) \, ,
\end{equation}
where $G(q)$ is the dressed quark propagator at zero chemical
potential. It has been shown that, with the renormalized D-S
equation, such an expression is equivalent to the one deduced from
the mass function in QCD Sum rules\cite{Craig04}. In the case of
finite chemical potential, one usually takes the same expression
to study the chemical potential dependence of the chiral quark
condensate\cite{MRS98,LGG03,Craig04}. We have then
\begin{equation}  \label{Qcondfmu}
|\langle\bar{q}q\rangle_{\mu}| = Tr_{D,C} \int
\frac{d^{4}q}{(2\pi)^{4}}\mathcal{G}[\mu](q) \, ,
\end{equation}
where ${\mathcal{G}}[\mu](q)$ is the dressed quark propagator at
finite chemical potential. It is then imperative to investigate the
chemical potential dependence of the dressed quark propagator.

The main aim of D-S equation approach is just to study the
propagators of fermion and boson at all energy scale. The D-S
equation approach is based on a coupled set of integral equations
among quark, gluon, ghost and vertex functions. They form a
countably infinite set of coupled integral equations with the one
for a $n$-point Schwinger function depending on the $(n+1)$ and
higher point functions. In order to handle the system practically,
it is necessary to make certain simplifications and truncations. One
simple approach, which is commonly referred to as the rainbow
approximation, employs the bare quark-gluon vertex and leaves the
equation for the dressed quark propagator to be solved with a given
effective gluon propagator as input. Under the rainbow 
approximation, the dressed quark propagator at zero chemical
potential $G(p)\equiv\mathcal{G}[\mu=0](p)$ can be determined well
with the truncated D-S equation\cite{Alkofer03,Alkofer04,Craig05}
\begin{equation}   \label{DSe0mu}
G^{-1}(p)=i\gamma\cdot p + \frac{4}{3}\int\frac{d^{4}q}{(2\pi)^{4}}
D_{\mu\nu}(p-q)\gamma_{\mu}G(q)\gamma_{\nu}   \, .
\end{equation}
where $D_{\mu\nu}(k)$ is an effective gluon propagator, and the
inverse of the dressed quark propagator is conventionally decomposed
as
\begin{equation}  \label{DecomAB}
G^{-1}(p)=i\gamma\cdot p A(p^2) + B(p^2)  \, ,
\end{equation}
with $A(p^2)$ and $B(p^2)$ being scalar functions. The chiral
quark condensate at zero chemical potential can be determined by
the two scalar functions as:
\begin{equation}\label{zero-cond}
|\langle\bar{q}q\rangle|=12\int
\frac{d^{4}q}{(2\pi)^{4}}\frac{B(q^2)}{q^{2}A^{2}(q^2)+B^{2}(q^2)}\,
.
\end{equation}

Basing on the solution of the D-S equation, one can obtain a
trivial solution $B(p^{2})=0$, namely the Wigner solution that
characterizes a phase in which chiral symmetry is not broken and
the dressed quarks are not confined. This phase relates to the
trivial vacuum with $\langle \bar{q} q \rangle \equiv 0$. We are
mainly interested in the nontrivial solution with nonzero
$B$(Nambu solution) which should be found by numerical iterations.
This solution corresponds to the nontrivial vacuum containing the
dynamical quark mass generation, the chiral quark condensate and
the existence of massless Goldstone bosons. It is also necessary
to mention that the Wigner solution is always possible due to the
path-integral formula used in the framework of the D-S equation.
Nevertheless, the Nambu solution is only possible if the coupling
is strong enough at the infrared
region\cite{Craig01,Pennington04,YCL06}. Since the chiral quark
condensate is zero in Wigner phase and nonzero in Nambu phase, the
variation behavior of the chiral quark condensate with respect to
the chemical potential can simulate the chiral phase transition
from the Nambu phase to Wigner phase.

The chemical potential is introduced as a Lagrange multiplier
$\mu\bar{q}\gamma_{4}q$ in QCD action. Similarly, the dressed
quark propagator ${\mathcal{G}}[\mu](p)$ at non-zero chemical
potential $\mu$ can be written as
\begin{equation}   \label{DSefmu}
\mathcal{G}^{-1}[\mu](p)=i\gamma\cdot
p-\gamma_{4}\mu+\frac{4}{3}\int\frac{d^{4}q}{(2\pi)^{4}}
D_{\mu\nu}(p-q)\gamma_{\mu}\mathcal{G}[\mu](q)\gamma_{\nu}  \, ,
\end{equation}
where the form of the effective interaction $D_{\mu\nu}$ is usually
taken as the same as that in Eq.~(\ref{DSe0mu}).

In prevenient approach of D-S equation at finite chemical
potential, the dressed-quark propagator can be written, in
general, as\cite{MRS98,Harada99}
\begin{equation}\label{DecomABC}
\mathcal{G}^{-1}[\mu](p)=i\vec{\gamma}\cdot\vec{p}A(\tilde{p}) +
i\gamma_{4}(p_{4}+i\mu)C(\tilde{p})+B(\tilde{p})  \, ,
\end{equation}
where $\tilde{p}_{\nu}=(\vec{p},p_{4}+i\mu)$, the complex functions
$A$, $C$ and $B$ can be defined by the quark equation with an
effective gluon propagator. With a $\delta$-function in momentum
space for the effective gluon propagator (Munczek-Nemirovsky
model~\cite{MN83}), it has been found that, in the D-S equation
approach~\cite{MRS98,LGG03}, the quark condensate increases with
chemical potential, which is not consistent with the behavior given
in composite operator approach of QCD~\cite{BC90},    
dilute instanton liquid model~\cite{CD99}, Nanbu--Jona-Lasinio (NJL)
model~\cite{Choe02} and other
approaches~\cite{CFG92,Cel95,GYL01,Druk03}. Looking over the
characteristic of the Munczek-Nomirovsky model, which gives an
infinity at zero exchanged momentum, we propose that the discrepancy
of the variation behaviors of the chiral quark condensate against
the chemical potential arises from the overestimated sudden
enhancement in the far infrared region. To remove the discrepancy,
we should then suppress the far infrared enhancement. Prevenient
works have shown that the model with effective gluon propagator
\begin{equation}   \label{Propg}
D_{\mu\nu}(k) = t_{\mu\nu}4\pi^{2}d\frac{\chi^{2}}{k^{4}+\Delta} \,
,
\end{equation}
with $d=\frac{12}{27}$ and $t_{\mu\nu}=\delta_{\mu\nu}$, which means
that the ``Feynman-like" gauge is taken in practical calculation,
can describe the pion weak decay constant and other low energy
chiral observables well (see for example Ref.\cite{Meis97}). It is
obvious that this effective gluon propagator has an finite infrared
enhancement and does not involve the ultraviolet behavior of the QCD
running coupling. Since such a model produces ultraviolet convergent
integrals naturally, the renormalization is not necessary.
Meanwhile, this effective gluon propagator does not have a Lehmann
representation, and the corresponding classical potential between
quarks can be written as
\begin{equation} \label{VPm4}
V(r)=-\frac{d \pi
\chi^{2}}{r\sqrt{\Delta}}e^{-r\sqrt[4]{\frac{\Delta}{4}}}
\sin[r\sqrt[4]{\frac{\Delta}{4}}] \, .
\end{equation}
It is apparent that such a potential corresponds to an
approximately quadratic confinement in intermediate range (up to
$r \approx 3.0$~fm) and linear-like in very short range ($r
\lesssim 0.2$~fm). In most recent years, great progress has been
made in systematic studies on quark-quark interaction (see for
example
Ref.~\cite{Alkofer03,Alkofer04,Zwanziger02,Smekel02,Watson04,Bowman04,Sternbeck05}).
One should, in principle, take the obtained analytic properties of
the gluon and quark propagators (c.f. given in
Ref.~\cite{Alkofer03,Alkofer04}) to study the chiral phase
transition. However, for simplicity in studying the quark D-S
equation and incorporating the chemical potential effect at
present stage, we take the effective gluon propagator in
Eq.~(\ref{Propg}) as the starting point of our model of
interaction.

In general approach to study the quark equation at finite chemical
potential, one should take a chemical potential $\mu$ dependent
effective gluon propagator as input, since the chemical potential of
the matter also influences the interaction. However, the related
knowledge is still in lack up to now. For simplicity, we propose an
approximation for the effective gluon propagator at finite chemical
potential with an extension $k^{2} \rightarrow k^{2} + \beta \mu
^{2}$,  the Eq.~(\ref{Propg}) is then modified as
\begin{equation}  \label{Propgmudp}
D_{\mu\nu}'(k) = t_{\mu\nu}4\pi^{2}d \frac{\chi^{2}}{(k^{2} + \beta
\mu^{2})^{2}+\Delta} \, ,
\end{equation}
where $\mu$ denotes the chemical potential and $\beta$ is a
scaling parameter denoting the strength of the $\mu$ dependence.
The classical potential with
this modified gluon propagator can be written as
\begin{equation}    \label{VPm4mudp}
V'(r)=-\frac{d \pi \chi^{2}}{r\sqrt{\Delta}}
e^{-r\sqrt{\frac{\sqrt{\beta^{2}\mu^{4} + \Delta}+\beta\mu^{2}}{2}}}
\sin[r\sqrt{\frac{\sqrt{\beta^{2}\mu^{4} +
\Delta}-\beta\mu^{2}}{2}}] \, .
\end{equation}
It is evident that the strength of the attractive interaction
becomes smaller with an increase of the chemical potential and the
scaling parameter $\beta$.

Substituting Eq.~(\ref{Propgmudp}) into Eq.~(\ref{DSefmu}) and
accomplishing some derivations, we have the equations for the scalar
functions $A(\tilde{p})$, $B(\tilde{p})$ and $C(\tilde{p})$ in
Eq.~(\ref{DecomABC}) as
\begin{equation}\label{DSEfmuA}
(A(\tilde{p})-1)\vec{p}^{2}=\frac{8}{3}\int\frac{d^{4}q}{(2\pi)^{4}}
\frac{4\pi^{2}\chi^{2} d }{(k^{2} + \beta\mu^{2})^{2} +
\Delta}\frac{A(\tilde{q})\vec{p}\cdot\vec{q}}{\vec{q}^{2}A^{2}(\tilde{q})
+(q_{4}+i\mu)^{2}C^{2}(\tilde{q})+B^{2}(\tilde{q})} \, ,
\end{equation}

\begin{equation}\label{DSEfmuC}
(C(\tilde{p})\!-\!1)(p_{4}+i\mu)=\frac{8}{3}\int\frac{d^{4}q}{(2\pi)^{4}}
\frac{4\pi^{2} \chi^{2} d }{(k^{2}\! + \!\beta \mu^{2})^{2}\! + \!
\Delta}\frac{C(\tilde{q})(q_{4}+i\mu)}{\vec{q}^{2}A^{2}(\tilde{q})
\! + \! (q_{4}\!+\!i\mu)^{2}C^{2}(\tilde{q})\!+\!
B^{2}(\tilde{q})} \, ,
\end{equation}

\begin{equation}\label{DSEfmuB}
B(\tilde{p})=\frac{16}{3}\int\frac{d^{4}q}{(2\pi)^{4}}
\frac{4\pi^{2} \chi^{2} d}{(k^{2} + \beta \mu^{2})^{2} +
\Delta}\frac{B(\tilde{q})}{\vec{q}^{2}A^{2}(\tilde{q}) +
(q_{4}+i\mu)^{2}C^{2}(\tilde{q})+B^{2}(\tilde{q})} \, .
\end{equation}

\noindent With the parameter being taken as
$\Delta=0.01~\mbox{GeV}^{4}$, $\chi=1.33$~GeV, with which the pion
decay constant can be fitted to $87$~MeV (quite close to the
experimental date $93$~MeV)\cite{Meis97}, we can solve the coupled
equations at zero chemical potential easily and obtain the chiral
quark condensate in vacuum as $(250~\mbox{MeV})^{3}$. It is
evident that such a result agrees excellently well with the
empirical value.

At nonzero chemical potential, it is a quite hard task to solve the
coupled integral equations (\ref{DSEfmuA})-(\ref{DSEfmuB}), since
they depend on the momenta $p$, the chemical potential $\mu$ as well
as the angle $\theta$ between $p$ and $\mu$ with $\cos{\theta}
=p\cdot\mu/\sqrt{p^{2}\mu^{2}}$ due to the breaking of the O(4)
symmetry in the four momentum space. By abandoning the commonly used
discretion of the unknown functions but implementing the
\emph{smooth polynomial approximations}\cite{Watson01,Bloch95} for
the to be determined complex functions $A(\tilde{p})$,
$C(\tilde{p})$ and $B(\tilde{p})$, we solve the equations
(\ref{DSEfmuA})-(\ref{DSEfmuB}) (for the process of the smoothing
and solving the equation, please see the Appendix-A).

\begin{figure}[hbtp]
\begin{center}
\includegraphics[scale=1,angle=0]{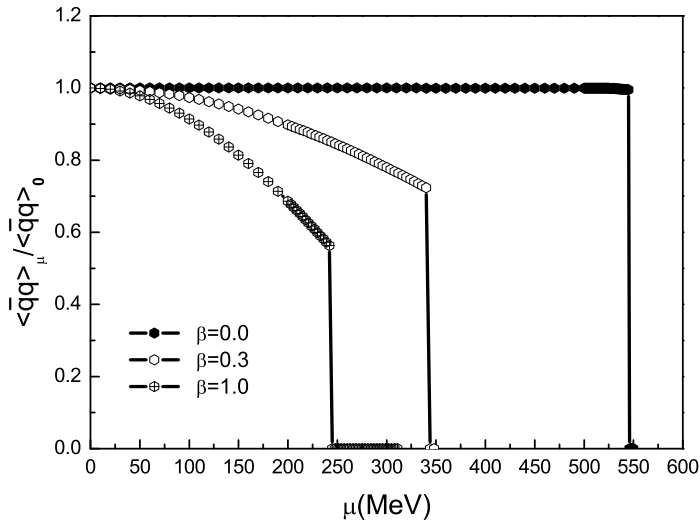}
\vspace*{0mm} \caption{\label{Qcond-mu} Calculated chemical
potential dependence of the chiral quark condensate at several
scaling parameters}
\end{center}
\end{figure}

\begin{figure}[hbtp]
\begin{center}
\includegraphics[scale=1,angle=0]{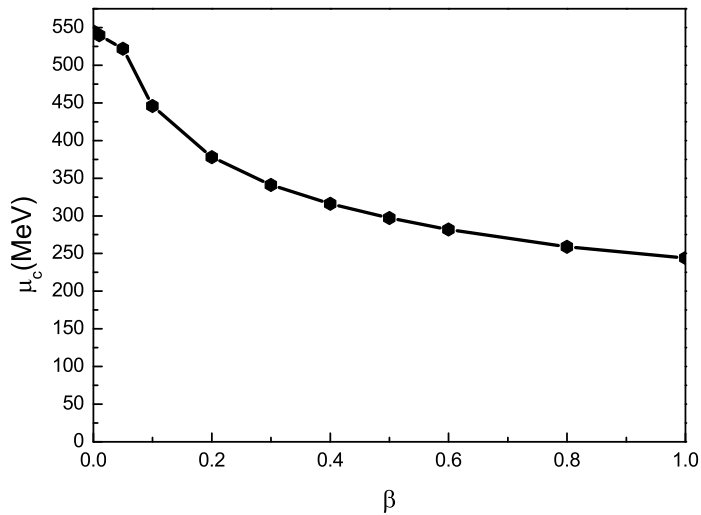}
\vspace*{0mm}\caption{\label{critmu-beta} The scaling parameter
dependence of the critical chemical potential for the chiral
symmetry to be restored}
\end{center}
\end{figure}

With the solutions of the D-S equations at any chemical potential
and scaling parameter $\beta$, we evaluate the chemical potential
dependence of the chiral quark condensate at a series values of the
scaling parameter $\beta$. A part of the obtained results are
illustrated in Fig.~{\ref{Qcond-mu}}.
The figure shows apparently that, as the chemical potential
increases in the region less than a critical value, the in-medium
chiral quark condensate decreases monotonously for any value of
$\beta \in (0, 1]$. As the critical value of the chemical
potential $\mu_{c}$ is reached, the chiral quark condensate
vanishes suddenly. Such a behavior indicates that the chiral
symmetry can be restored partially before the critical chemical
potential is reached. At the critical point of the chemical
potential, the chiral symmetry can be completely restored
suddenly. And the phase transition is in first order. Such a
result is consistent with that given in the composite operator
formalism of QCD~\cite{BC90}. Moreover, with the increase of the
scaling parameter $\beta$, the decreasing rate of the chiral quark
condensate against the chemical potential gets obviously larger
and the critical chemical potential becomes apparently smaller.
For instance, for $\beta=0.3$, $1.0$, the critical chemical
potential $\mu_{c}$ is $344$~MeV, $245$~MeV and the ratio
$|\langle\bar{q}q\rangle_{\mu_{c}}|/|\langle\bar{q}q\rangle_{0}|$
is $72.4\%$, $56.3\%$, respectively. In addition, in the case of
that the chemical potential does not affect the effective gluon
propagator, i.e., $\beta \equiv 0$, the chiral quark condensate in
the matter maintains almost a constant if the chemical potential
is less than the critical value ($\mu_{c} = 546$~MeV,
$|\langle\bar{q}q\rangle_{\mu_{c}}|/|\langle\bar{q}q\rangle_{0}|=99.6\%$).
Such a result is analogous to that given in the NJL
model~\cite{Choe02} and dilute instanton liquid model~\cite{CD99}.
These results indicate that the critical chemical potential for
the chiral quark condensate to be restored completely decreases
with the increase of the scaling parameter $\beta$. The obtained
result of the scaling parameter dependence of the critical
chemical potential is illustrated in Fig.~{\ref{critmu-beta}}.

\begin{figure}[hbtp]
\begin{center}
\includegraphics[scale=1,angle=0]{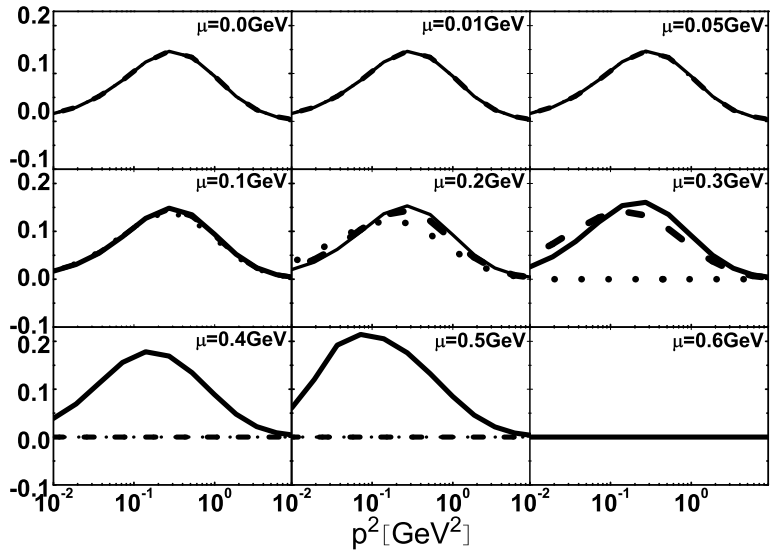}
\vspace*{0mm} \caption{\label{IGpmu-beta} Calculated chemical
potential dependence of $IG(p^{2})$ at several scaling parameters:
solid curve for $\beta=0.0$; dash curve for $\beta=0.3$; dot curve
for $\beta=1.0$.}
\end{center}
\end{figure}

To understand the variation characteristics of the chiral quark
condensate against the chemical potential and the scaling parameter
intuitively, we recall the definition of the condensate in
Eq.~(\ref{Qcondfmu}). It is evident that the integrand for the
condensate reads
\begin{equation}\label{integ}
\frac{p^{2} B(\tilde{p})}{\vec{p}^{2}A^{2}(\tilde{p}) +
(p_{4}+i\mu)^{2}C^{2}(\tilde{p})+B^{2}(\tilde{p})} \, ,
\end{equation}
where the $A(\tilde{p})$, $C(\tilde{p})$ and $B(\tilde{p})$ are
the solutions of the D-S equations
(\ref{DSEfmuA})-(\ref{DSEfmuB}). After integrating over the angle,
one can express the momentum dependence of the integrand
explicitly as (neglecting some constants)
\begin{equation}\label{IGG}
IG(p^{2})=\int_{0}^{\pi}d\theta \sin^{2}\theta\frac{ p^{2}
B(\tilde{p})}{\vec{p}^{2}A^{2}(\tilde{p}) +
(p_{4}+i\mu)^{2}C^{2}(\tilde{p})+B^{2}(\tilde{p})} \, ,
\end{equation}
with $\cos{\theta} = p\cdot\mu/\sqrt{p^{2}\mu^{2}}$. Since the real
part of both $A(\tilde{p})$ and $C(\tilde{p})$ is even and
increasing function of the $\cos{\theta}$, the real part of
$B(\tilde{p})$ is an even but decreasing function of the
$\cos{\theta}$, and the imaginary part of $A(\tilde{p})$,
$C(\tilde{p})$ and $B(\tilde{p})$ is odd function of $\cos{\theta}$,
such an integrand is a real function of the momentum $p$. The
behavior of the function $IG(p^{2})$ at several values of the
chemical potential and the scaling parameter can be displayed in
Fig.~{\ref{IGpmu-beta}}.
It is evident that, at any chemical potential, the function
$IG(p^{2})$ is convergent in the infrared and the ultraviolet
regions. It indicates that the condensate is well defined. In more
detail, there exists a critical chemical potential $\mu_{c}$, if
$\mu > \mu_{c}$, $IG(p^{2}) \equiv 0 $. Such a $\mu_{c}$ decreases
with the increasing of the scaling parameter $\beta$, for example,
$\mu _{c} = 0.546$~GeV for $\beta=0.0$, $\mu _{c} = 0.344$~GeV for
$\beta=0.3$, $\mu_{c} = 0.245$~GeV for $\beta=1.0$. In the case of
$\beta \ne 0$, as $\mu < \mu_{c}$,
$IG(p^{2})$ increases with the increasing of momentum. However the
$IG(p^{2})$ decreases at large momentum. The momentum for the
$IG(p^{2})$ to take the maximal value shifts to smaller as the
chemical potential and the scaling parameter increase. Moreover,
the maximal value of $IG(p^{2})$ decreases as the scaling
parameter increases. These characteristics of the integrand
manifests evidently that, if the chemical potential is smaller
than a critical value but influences the effective gluon
propagator (i.e., with $\beta \in (0,1]$), the chiral quark
condensate decreases with the increasing of chemical potential,
and the decreasing rate gets larger as the scaling parameter
increases. As the chemical potential reaches the critical value,
the chiral quark condensate vanishes suddenly. In the case of that
the chemical potential does not influence the effective gluon
propagator (i.e., with $\beta \equiv 0$), even though the
solutions of the D-S equation at finite chemical potential are
different from those at zero chemical potential, because of the
compensation of the solutions $A$, $C$ and $B$ at the presently
chosen parameters, the integrand of the chiral quark condensate
behaves almost independent of the chemical potential if it is less
than the critical value. The chiral quark condensate turns to be
then approximately a constant before it vanishes suddenly at the
critical chemical potential. On the other hand, concerning the
mass function of a quark in the case of that the chemical
potential does not affect the effective gluon propagator, we
realize that it does not change drastically either with the
chemical potential as it is smaller than the critical value. It
means that the dressing effect come from the chiral quark
condensate does not change obviously. It hints then the chiral
quark condensate maintains almost a constant if the effective
gluon propagator is independent of the chemical potential.

To explore the underlying physics of the variation behavior of the
chiral quark condensate against the chemical potential and the
scaling parameter, we look through the characteristic of the model
effective gluon propagator. From Eqs.~(\ref{Propgmudp}) and
(\ref{VPm4mudp}), one can recognize easily that the model
effective gluon propagator is the one with approximately quadratic
confinement in intermediate region and linear-like in very short
region. Both the chemical potential and the scaling parameter play
roles of screening on the confinement. It means that the increases
of the chemical potential and the scaling parameter decrease the
confinement strength. It has been known that the chiral quark
condensate is proportional to the mean spectral density at zero
energy of quarks with Banks-Casher relation\cite{Craig04,BC80}
$-\langle \bar{q} q \rangle = \pi \nu_{q}(0)$. In general
principle, the mean spectral density $\nu(0)$ decreases as the
attractive interaction gets weaker. The classical potential in
Eq.~(\ref{VPm4mudp}) shows that, in the case of that the effective
gluon propagator depends on the chemical potential (i.e., with
scaling parameter $\beta > 0$), the interaction in the system is
just attractive and gets weaker as the chemical potential and the
scaling parameter increase. Corresponding to the increases of the
chemical potential and the scaling parameter, the mean spectral
density at zero energy of quarks decreases, so that the absolute
value of the chiral quark condensate decreases. Moreover, as the
chemical potential is large enough (i.e., the density of quarks is
sufficiently large), the screening effect is so strong that the
interaction between quarks becomes very weak, even vanishes
(asymptotic free), which is analogous to the molecular in a
liquid, where it experiences an attraction to the inner if it is
in the surface region and becomes almost free in the inner region.
As a consequence, the chiral quark condensate changes to zero
suddenly. Combining the calculated variation behavior of the
chiral quark condensate with respect to the chemical potential (or
the density of the strong interaction matter) and the inferred
changing characteristic of the interaction between quarks, we can
recognize that the chiral symmetry can be restored abruptly as the
density of the matter reaches a critical value, and the chiral
phase transition is in first order. In addition, the chiral
symmetry can be restored partially and gradually before the phase
transition takes place. However, in the case of that the effective
gluon propagator is independent of the chemical potential (i.e.,
with $\beta \equiv 0$), the interaction is free from the chemical
potential, or in other word, the chemical potential does not
screen the attractive interaction. Then the chiral quark
condensate almost maintains a constant before the critical
chemical potential is reached.


In summary, we have calculated the chiral quark condensates in
vacuum and in strongly interacting matter in the Dyson-Schwinger
equation approach with a model effective gluon propagator
including finite chemical potential effect in this paper. The
calculated results show that, as the chemical potential influences
the effective gluon propagator, the chiral quark condensate
decreases with the increase of the chemical potential before the
critical value is reached and the condensate vanishes suddenly at
the critical chemical potential. Moreover, the increase of the
scaling parameter of the chemical potential in the effective gluon
propagator enhances the decreasing rate and decreases the critical
chemical potential. It indicates that the chiral symmetry can be
restored abruptly as the density of the matter reaches a critical
value, and the chiral phase transition is in first order. Before
the critical density or chemical potential is reached, the chiral
symmetry can be restored partially and gradually. In addition, as
the chemical potential does not affect the effective gluon
propagator, the chiral symmetry can only be restored suddenly at a
critical chemical potential, but no gradual restoration takes
place before the critical chemical potential is reached. It
manifests that the effect of the chemical potential on the gluon
propagator plays significant role in the process of the chiral
symmetry restoration. Concerning the modification on the effective
gluon propagator at finite chemical potential, we just extend the
exchanged momentum related term from $k^{2}$ to chemical potential
dependent $k^{2} + \beta \mu ^{2}$. The parameter $\beta$ or the
$\beta \mu ^{2}$ displays the effect of the finite chemical
potential on the effective gluon propagator or the quark-quark
interaction. This is definitely only a model or an approximation.
Furthermore, we have taken the rainbow approximation for the quark
equation. Then studying the strong interaction matter by solving
the coupled quark, gluon and ghost equations systematically is
imperative. The related investigations are under progress.

\bigskip

\bigskip

This work was supported by the National Natural Science Foundation
of China (NSFC) under contract Nos. 10425521 and 10575004, the Major
State Basic Research Development Program under contract No.
G2000077400, the research foundation of the Ministry of Education,
China (MOEC), under contact No. 305001 and the Research Fund for the
Doctoral Program of Higher Education of China under grant No
20040001010. One of the authors (YXL) thanks also the support of the
Foundation for University Key Teacher by the MOEC.

\bigskip

\noindent{\bf Appendix-A:}

Because the O(4) symmetry in the four momentum is broken, the
complex functions $A(\tilde{p})$, $C(\tilde{p})$ and $B(\tilde{p})$
depend not only on the square of momentum $p^{2}$ and chemical
potential $\mu^{2}$ but also on the angle $\theta$ between $p$ and
$\mu$ with $ \cos{\theta}=p\cdot\mu/\sqrt{p^{2}\mu^{2}}$. To solve
the coupled equations (\ref{DSEfmuA})-(\ref{DSEfmuB}), a smooth
polynomial approximation should be taken for both the angle and the
square of momentum when we discrete the integrand in numerical
calculation.

The angle $\theta$ dependence of the functions is analyzed by an
expansion in terms of the second kind Chebyshev polynomials
$U_{i}(\cos{\theta})$ as
$$\displaylines{\hspace*{1cm}
A(\tilde{p}^{2})=\sum_{m=0}^{N_{ch,2}-1}A_{m}(p^{2};\mu^{2})U_{m}(\cos{\theta})(i)^{m},
\hfill{(A.1)} \cr \hspace*{1cm}
C(\tilde{p}^{2})=\sum_{m=0}^{N_{ch,2}-1}C_{m}(p^{2};\mu^{2})U_{m}(\cos{\theta})(i)^{m},
\hfill{(A.2)} \cr \hspace*{1cm}
B(\tilde{p}^{2})=\sum_{m=0}^{N_{ch,2}-1}B_{m}(p^{2};\mu^{2})U_{m}(\cos{\theta})(i)^{m}.
\hfill{(A.3)} \cr }
$$
Since the $U_{k}(\cos{\theta})$ form an orthogonal set, we can
project the equations (\ref{DSEfmuA}), (\ref{DSEfmuC}) and
(\ref{DSEfmuB}) onto a set of nonlinear integral equations
satisfied by the form factors $A_{m}(p^{2};\mu^{2})$,
$C_{m}(p^{2};\mu^{2})$ and $B_{m}(p^{2};\mu^{2})$ which are all
\emph{real} functions depending only on $p^{2}$ and $\mu^{2}$. In
practical calculation we truncate the expansion at a certain rank
$N_{ch,2}$ and the degree of the fitting can be mitigated by
increasing $N_{ch,2}$. After the angular expansion we then take
Chebyshev expansion for the form factors $A_{m}(p^{2};\mu^{2})$,
$C_{m}(p^{2};\mu^{2})$ and $B_{m}(p^{2};\mu^{2})$, which gives
$$\displaylines{\hspace*{1cm}
A_{m}(p^{2};\mu^{2})=\sum_{j=1}^{N_{ch,1}}a_{m}^{j}T_{j}(s(p))-\frac{a_{m}^{1}}{2},
\hfill{(A.4)} \cr \hspace*{1cm}
C_{m}(p^{2};\mu^{2})=\sum_{j=1}^{N_{ch,1}}c_{m}^{j}T_{j}(s(p))-\frac{c_{m}^{1}}{2},
\hfill{(A.5)} \cr \hspace*{1cm}
B_{m}(p^{2};\mu^{2})=\sum_{j=1}^{N_{ch,1}}b_{m}^{j}T_{j}(s(p))-\frac{b_{m}^{1}}{2},
\hfill{(A.6)} \cr }
$$
 with
$$s(p)=\frac{\log(p^{2}/\Lambda\epsilon)}{\log(\Lambda/\epsilon)}
\, , $$
 where $T_{j}(s(p))$ is the $j$-rank Chebyshev polynomial
of the first kind, $\Lambda$ is the ultraviolet cutoff, and
$\epsilon$ is the infrared cutoff. In this expansion we make use
of the first kind Chebyshev polynomial with a cutoff $N_{ch,1}$.
Up to now the unknown functions $A$, $C$ and $B$ have been
expressed by the smooth polynomial with $3 N_{ch,1} N_{ch,2}$
Chebyshev coefficients $a_{m}^{j}$, $c_{m}^{j}$ and $b_{m}^{j}$
which would be numerically calculated by an effective iteration
method. In practical calculation, we take $N_{ch,1}=32$,
$N_{ch,2}=6$, $\Lambda=10$~GeV and $\epsilon=0.01$~GeV
(considering the parameter $\Delta = 0.01\mbox{GeV}^{4}$ and
$\epsilon^{4} = 10^{-8}\mbox{GeV}^{4}$, such a choice of the
cutoff is quite reasonable).

In order to determine the $3 N_{ch,1} N_{ch,2}$ Chebyshev
coefficients $a_{m}^{j}$, $c_{m}^{j}$ and $b_{m}^{j}$, we require
these integral equations to satisfy $N_{ch,1}$ fixed external
momenta. To solve the nonlinear equations, we implement the Newton
method. This method takes derivatives of the equations with
respect to the unknowns to speed up the convergence. After some
derivations, it is transformed to a set of ($(3 N_{ch,1}
N_{ch,2})\times (3 N_{ch,1} N_{ch,2})$) linear equations, which
can be solved easily. As for the integration we introduce
spherical coordinates, i.e. the volume element $d^{4}q$ reads
$2\pi q^{3}dq \sin^{2}\theta d\theta \sin\phi d\phi$. The
integration ranges of the variables are: $q\in [\epsilon,\Lambda]$
and $\theta,\phi \in [0,\pi]$. The integration is calculated by
the N-points Gauss-Legendre quadrature rule. The integrand may
still not be smooth enough at the junctions of some intervals, it
is then necessary to split the region for much higher accuracy. We
split the integrand into three regions
$[\epsilon^{2},min(p^{2},q_{f}^{2})]$,
$[min(p^{2},q_{f}^{2}),max(p^{2},q_{f}^{2})]$ and
$[max(p^{2},q_{f}^{2}),\Lambda^{2}]$, where $p^{2}$ is the square
of the external momentum and $q_{f}^{2}$ is defined by the
possible zero of $\vec{q}^{2}A^{2}(\tilde{q}) +
(q_{4}+i\mu)^{2}C^{2}(\tilde{q})+B^{2}(\tilde{q})$ at $q_{4}=0$.




\begin{thebibliography}{50}
\bibitem{SS98} T. Sch\"{a}fer and E. V. Shuryak, Rev. Mod. Phys.
{\bf 70} (1998), 323; and references therein.
\bibitem{BR02} G. E. Brown and M. Rho, Phys. Rept. {\bf 363} (2002), 85.
\bibitem{MST04}  \'{A}. M\'{o}csy, F. Sannino, and K.
Tuominen, Phys. Rev. Lett. {\bf 92} (2004), 182302; and references
therein.
\bibitem{Choe01} S. Choe {\it et al.}, QCD-TARO Collaboration, Nucl.
Phys. {\bf B}(Proc. Suppl.) {\bf 106} (2002), 462.
\bibitem{FK02} Z. Fodor, and S. D. Katz, Phys. Lett. {\bf B 534}
(2002), 87.
\bibitem{Allton02} C.R. Allton, S. Ejiri, S.J. Hands, O.
Kaczmarek, F. Karsch, E. Laermann, Ch. Schmidt, and L. Scorzato,
Phys. Rev. {\bf D 66} (2002), 074507.
\bibitem{Delia03} M. D'Elia, and M.P. Lombardo, Phys. Rev. {\bf D
67} (2003), 014505.
\bibitem{GG03} R.V. Gavai, and S. Gupta, Phys. Rev. {\bf D 68}
(2003), 034506.
\bibitem{Focrand} P. de Focrand, and O. Philipsen, Nucl. Phys.
{\bf B 642} (2002), 290; {\it ibid}, {\bf B 673} (2003), 170; S.
Kratochvila, and P. de Focrand, arXiv: hep-lat/0409072; arXiv:
hep-lat/0509143; O. Philipsen, arXiv: hep-lat/0510077; P. de
Focrand, and S. Kratochvila, Nucl. Phys. {\bf B} (Proc. Suppl.) {\bf
153} (2006), 62.
\bibitem{ADGL05} V. Azcoiti, G. Di Carlo, A. Galante, V. Laliena,
Nucl. Phys. {\bf B 723} (2005), 77.
\bibitem{AFHL05} A. Alexandru, M. Faber, I. Hovath, and K.F. Liu,
Phys. Rev. {\bf D 72} (2005), 114513.
\bibitem{KL99} F. Karsch and M. L\"{u}tgemeier, Nucl. Phys. {\bf B 550}
 (1999), 449.

\bibitem{BC90} A. Barducci, R. Casalbuoni, S. De Curtis, R. Gatto,
G. Pettini, Phys. Rev. {\bf D 41} (1990), 1610.
\bibitem{CFG92} T. D. Cohen, R. J. Furnstahl and D. K. Griegel,
Phys. Rev. {\bf C 45} (1992), 1881.
\bibitem{LK94} G. Q. Li and C. M. Ko, Phys. Lett. {\bf B 338} (1994),
118.
\bibitem{Cel95} L. S. Celenza, C. M. Shakin, W. D. Sun, and J. Szweda,
Phys. Rev. {\bf C 51} (1995), 3372.
\bibitem{AB01}A. Bender, D. Blaschke, Yu. Kalinovsky, and C. D.
Roberts, Phys. Rev. Lett. {\bf 77} (1996), 3724.
\bibitem{MD97} M. Malheiro, M. Dey, A. Delfino and J. Dey, Phys.
Rev. {\bf C 55} (1997), 521.
\bibitem{Mits97} T. Mitsumori, N. Noda, H. Kouno, A. Hasegawa and
M. Nakano, Phys. Rev. {\bf C 55} (1997), 1577.
\bibitem{MRS98} P. Maris, C. D. Roberts, S. Schmidt, Phys. Rev. {\bf C
57} (1998), R2821.
\bibitem{CD99} G. W. Carter, and D. Diakonov, Phys. Rev. {\bf D
60} (1999), 016004.
\bibitem{GYL01} H. Guo, S. Yang anf Y.X. Liu, Sci. in China {\bf A
45} (2001), 334.
\bibitem{Choe02} O. Miyamura, S. Choe, Y. Liu, T. Takaishi, and
A. Nakamura, Phys. Rev. {\bf D 66} (2002), 077502.
\bibitem{LGG03} Y. X. Liu, D. F. Gao, and H. Guo, Phys. Rev.
{\bf C 68} (2003), 035204.
\bibitem{Druk03} E. G. Drukarev, Prog. Part. Nucl. Phys. {\bf 50} (2003),
659; and references therein.
\bibitem{Zong05} Hong-shi Zong, Lei Chang, Feng-yao Hou,
Wei-min Sun, and Yu-xin Liu, Phys. Rev. {\bf C 71} (2005), 015205.
\bibitem{Craig01} C. D. Roberts and A. G. Williams, Prog. Part.
Nucl. Phys. {\bf 33} (1994), 477.
\bibitem{Alkofer01} R. Alkofer, and L. von Smekal, Phy. Rept.
{\bf 353} (2001), 281; R. Alkofer, M. Kloker, A. Krassnigg, and R.
F. Wagenbrunn, Phys. Rev. Lett. {\bf 96} (2006), 022001.
\bibitem{Craig02} P. Maris and C. D. Roberts, Int. J. Mod. Phys.
{\bf E 12} (2003), 297; and references therein.
\bibitem{Watson01} R. Alkofer, P. Watson, and H. Weigel, Phy. Rev.
{\bf D 65} (2002), 094026.
\bibitem{Harada99} M. Harada, and A. Shibata, Phys. Rev. {\bf D 59}
(1999), 014010.
\bibitem{Craig03} C. D. Roberts, and S. M. Schmidt, Prog. Part.
Nucl. Phys. {\bf 45} (2000), S1; and references therein.

\bibitem{Craig04} K. Langfeld, H. Markum, R. Pullirsch, C. D. Roberts,
and S. M. Schmidt, Phy. Rev. {\bf C 67} (2003), 065206.
\bibitem{Alkofer03} C. S. Fischer, and R. Alkofer, Phy. Rev.
{\bf D 67} (2003), 094020.
\bibitem{Alkofer04} R. Alkofer, W. Detmold, C. S. Fischer, and P.
Maris, Phys. Rev. {\bf D 70} (2004), 014014.
\bibitem{Craig05} M. S. Bhagwat, M. A. Pichowsky, C. D. Roberts,
and P. C. Tandy, Phy. Rev. {\bf C 68} (2003), 015203.
\bibitem{Pennington04} M. R. Pennington, QCD Down Under: Building Bridges,
arXiv:hep-ph/0409156.
\bibitem{YCL06} W. Yuan, H. Chen, and Y. X. Liu, Phys. Lett. {\bf
B 637} (2006), 69.
\bibitem{MN83} H. J. Munczek, and A. M. Nemirovsky, Phys. Rev. {\bf D
28} (1983), 181.
\bibitem{Meis97} T. Meissner, Phys. Lett. {\bf B 405} (1997), 8.
\bibitem{Zwanziger02} D. Zwanziger, Phy. Rev. {\bf D 65} (2002), 094039.
\bibitem{Smekel02} C. Lerche, and L. von Smekel, Phy. Rev.
{\bf D 65} (2002), 125006.
\bibitem{Watson04} P. Watson, W. Cassing, and P. C. Tandy,
Few-Body Systems {\bf 35} (2004), 129.
\bibitem{Bowman04} P. O. Bowman, U. M. Heller, D. B. Leinweber,
M. B. Parappilly, and A. G. Williams, Phy. Rev. {\bf D 70} (2004),
034509; J.I. Skullerud, P.O. Bowman, A. Kizilersu, D.W. Leinweber,
A.G. Williams, Nucl. Phys. B (Proc. Suppl.) {\bf 141} (2005), 244;
P. O. Bowman, U. M. Heller, D. B. Leinweber, M. B. Parappilly, A. G.
Williams and J.B. Zhang, Phy. Rev. {\bf D 71} (2005), 054507.
\bibitem{Sternbeck05} A. Strenbeck, E.-M. Ilgenfritz, and
M. M\"{u}ller-Preussker, Phy. Rev. {\bf D 72} (2005), 014507.

\bibitem{Bloch95} J. C. Bloch, Ph. D. Thesis, University of Durham,
1995, arXiv:hep-ph/0208074.

\bibitem{BC80} T. Banks, and A. Casher, Nucl. Phys. {\bf B 169}
(1980), 103.

%

\end{thebibliography}
\end{document}